\documentclass[preprint,authoryear,12pt]{elsarticle} 
\usepackage[left=2cm,right=2cm,top=2cm,bottom=2cm]{geometry}

\usepackage[latin1]{inputenc}
\usepackage{natbib}
\usepackage{amssymb}
\usepackage{multicol}
\usepackage[fleqn]{amsmath}
\usepackage{epsfig}
\usepackage[normalem]{ulem}
\usepackage{verbatim}
\usepackage{graphicx}
\usepackage{color}
\usepackage{bbm}
\usepackage{bm}
\usepackage{dsfont}
\usepackage{ulem}

\usepackage{amssymb}

\usepackage{lineno}

\usepackage{verbatim}
\usepackage{graphicx}
\usepackage{color}
\usepackage{bbm}
\usepackage{bm}
\usepackage{dsfont}
\usepackage{amsmath}

\newcommand{\1}{\mathbbm{1}}

\newcommand{\Dir}{\mbox{Dir}}
\newcommand{\Be}{\mbox{Be}}

\begin{document}

\begin{frontmatter}



\title{Estimating Discrete Markov Models From Various Incomplete Data Schemes}

\author[EDFRD]{Alberto Pasanisi\corref{cor1}}
\ead{alberto.pasanisi@edf.fr}
\author[EDFRD,UP11]{Shuai Fu}
\ead{shuai.fu@edf.fr}
\author[EDFRD]{Nicolas Bousquet}
\ead{nicolas.bousquet@edf.fr}
\address[EDFRD]{EDF R\&D, Industrial Risk Management Department, 6 quai Watier, 78401 Chatou (France)}
\address[UP11]{University of Paris-Sud 11, Mathematics Department, Bat. 425, 91405 Orsay (France)}
\cortext[cor1]{Corresponding author. Tel: +33 1 30878085, Fax +33 1 30878213}

\begin{abstract}
The parameters of a discrete stationary Markov model are transition probabilities between states. Traditionally, data consist in sequences of observed states for a given number of individuals over the whole observation period. In such a case, the estimation of transition probabilities is straightforwardly made by counting one-step moves from a given state to another.
In many real-life problems, however, the inference is much more difficult as state sequences are not fully observed, namely the state of each individual is known only for some given values of the time variable.
A review of the problem is given, focusing on Monte Carlo Markov Chain (MCMC) algorithms to perform Bayesian inference and evaluate posterior distributions of the transition probabilities in this missing-data framework. Leaning on the dependence between the rows of the transition matrix, an adaptive MCMC mechanism accelerating the classical Metropolis-Hastings algorithm is then proposed and empirically studied.
\end{abstract}


\end{frontmatter}


\section{Introduction}\label{intro}

When facing situations where a variable of interest $Z$ takes time-dependent values within a finite (discrete) set ${\bf S} = \{ {s_1 ,s_2 ,\ldots,s_r }\}$ of $r$ classes (let us call them {\it states}), a decision-maker is most often interested in estimating the probability $p_{\cal{A}}(t)$ for $Z$ to be in a given set of states ${\cal{A}}\subset{\bf S}$ as a function of time $t$. For instance, these states can correspond to failure states in reliability analysis, thus  $1-p_{\cal{A}}(t)$ is the reliability function of the system under investigation $\Sigma$ at time $t$. Another function of interest could be the expected number of states $N_{{\cal{A}}}$ before $\Sigma$ reaches ${\cal{A}}$. To do so, the decision-maker first has to estimate the transition probabilities from one state to another, i.e. estimate the transition matrix $\bm\psi$. The vector $\bm{p}(0)$ of the initial probabilities $p_1(0), \dots p_r(0)$ for the system to be in states $s_1, \dots s_r$ respectively, at $t=0$, is usually assumed known in real-life applications; therefore, the knowledge of the transition matrix $\bm{\psi}$ allows to evaluate, for a given time $t$, the probabilities for being in each of the $r$ states, i.e. the vector:
\begin{eqnarray}
\bm{p}(t) = \bm{p}(0) \cdot {\bm{\psi}} ^t \label{probabilities.0}.
\end{eqnarray}
Given some data $\bm z$ under the form of observed sequences of states, the statistical estimation of these probabilities is traditionally facilitated by a time-homogeneous, first-order Markov stationarity assumption about the process $\bm\Lambda$ which generates the data. In other words, the transition probability $\psi_{i,j}$ from any state $s_i$ to any other state $s_j$ ($i$ possibly equals to $j$) is assumed to be independent of time and of the past trajectories before reaching $s_i$. As noted by \citet{JON05}, this assumption can seem somewhat restrictive regarding the external knowledge about the process, but ``using (possibly more appropriate) higher-order processes increases the complexity and data requirements quite substantially, and may not be feasible with only a limited time series'', which is often the case in practice. Such situations are encountered in a wide range of application domains, as we discuss now.

To start with, the applications of Markov models in engineering practice are numerous.
In reliability engineering and safety analysis, discrete Markov schemes, the states of which correspond to gradually degraded operating conditions, have for instance been used to assess the reliability of programmable electronic systems \citep{BUK95}, cogeneration plants \citep{ELN08}, machineries of oil refineries \citep{COC01},
 water meters \citep{PAS04}, piping systems of power plants \citep{CRO09}
 and welded structures submitted to fatigue damage \citep{LAS91}. In this  paper, the example treated in Section \ref{complete.exemple} provides a real industrial use-case where a Markov model is used to assess the reliability of rotating machines. 
Examples of application in hydrology and water resources engineering concern the modeling of river inflows \citep{PAR91}, lake inflows \citep{DUC79}, water supply reservoir states \citep{VOG87} or pollutants propagations \citep{ANG84, ZHA07}. 
In biomedical survey, Markov chains can model the health condition of patients affected by infectious or viral diseases like in \citet{GEN94}. These models are also applied to capture-recapture problems \citep{DUP95,DUP07}, used to describe the dynamics of an animal population. As a last example, the financial world makes a wide use of first-order Markov transition matrices to explain the migration of credit ratings \citep{JON05,FUE07} or to model loan defaults {\citep{GRI11}}.

In an ideal framework, the data ${\bm z}$ consist in $m$ time series of observed states for $m$ identical individuals (systems) $\Sigma$ that are assumed independent. If no data is missing, the estimation of $\bm\psi$ is relatively straightforward. 
In many applied problems, however, part of data is missing. Such problems can often be divided in two classes.
\begin{description}
\item {\bf (i)} We call an {\it incomplete sequence problem} the estimation of $\bm\psi$ when ${\bm z}$ are observed trajectories of states:
\begin{eqnarray*}
\begin{array}{ccccc}
z_{(1,1)} & \bullet & \dots  & z_{(1,T-1)} & z_{(1,T)}\\
\bullet & z_{(2,2)} & \dots  & z_{(2,T-1)} & z_{(2,T)}\\
\vdots & \vdots & \vdots &  \vdots  & \vdots\\
z_{(m,1)}& \bullet & \dots & z_{(m,T-1)} & \bullet,
\end{array}
\end{eqnarray*}
containing random missing items (random successions of unknown states symbolized by ``$\bullet$"), assuming the initial state is known. This occurs typically when the $m$ individuals are checked at deterministic times $t=1,\ldots,T$, independently from $\bm\Lambda$, as noted by \citet{DUP95} or when the survey of all individuals at the same time is impossible (e.g. only a given proportion of the machineries can be inspected simultaneously, in order to avoid stopping the industrial production). 
\item {\bf (ii)}  We call an {\it aggregate data problem} the estimation of $\bm\psi$ when the sequential data ${\bm z}$ are reduced to the numbers of individuals $n_i(t)$ being in a given state $s_i$ at a given time $t$ (i.e. $n_i(t)=\sum_{j=1}^m \1_{\{{z_{(j,t)}= s_i}\}}$). Such data are frequently  \citep{GOU11} the only ones being at disposal of the analyst, because, for instance, the full trajectories of individuals represented too much information or were not considered of primary importance during the survey process.
\end{description} 
Our first aim in this paper is to give in Section \ref{review} a short review of the main computational methods dedicated to the estimation task in the complete data scheme, as well as in both missing data schemes described above, in a Bayesian statistical framework that we defend as being the most convenient. Monte Carlo Markov Chains (MCMC) \citep{ROB04} are used to deal with such missing data schemes. More specifically, we address ourselves to the reader who is interested in reducing the computational cost, 
which appears as a practical difficulty when the number of degrees of freedom is high and/or when the Markov model is part of a larger, encompassing model.  Simulated experiments help us give some guidelines to select a variant of these methods in the common case where, for each individual, the state at just one time step has been observed.

In Section \ref{research} we consider the time-consuming aspects of numerical computation. Two adaptation mechanisms, which explore the correlations between the transition probabilities, are proposed to accelerate the convergence of the Metropolis-Hastings algorithm towards the posterior distribution. 
A series of numerical experiments based on a class of transition matrices largely encountered in reliability analysis is led in Section \ref{tests}. Although they remain relatively basic, we show that our proposals can help reduce the computation time significantly. 
Finally, a discussion section sums up the main results and advices arising from both main parts of the paper, and highlights several promising research avenues, especially in the theoretical description of adaptive MCMC.

\section{Bayesian estimation of transition probabilities: a review}\label{review}

This section provides a review of Bayesian inference techniques for the estimation of the transition matrix $\bm \psi$ under the obvious conditions:
\begin{eqnarray}
0\leq \psi_{i,j} \leq1, ~~~~\sum\limits_{j=1}^{r}{\psi_{i,j}}=1.
\label{transition.conditions}
\end{eqnarray}
This estimation problem has thus $r(r-1)$ degrees of freedom. As stated hereinbefore, we voluntarily chose a Bayesian viewpoint. Besides the more theoretical issues pointed by \citet{ROB01}, we motivate our choice, in an industrial context, by the possibility to explicitly (and relatively easily) quantify, via predictive simulation \citep{GIR04}, the uncertainty affecting some quantities of practical interest for the reliability engineer (e.g. the probability for the system to be in a failure state for a given  time $t$, or the mean time before the system reaches one of the failure states).
Moreover, from a strictly computational point of view, the Bayesian framework allows here to deal with some issues that can be quite burdensome in frequentist inference, without any particular additional difficulty. These include the intractability of the likelihood expression in missing data schemes, the respect of constraints (\ref{transition.conditions}), the difficulty to obtain a probability distribution for the estimators $\hat{\bm \psi},$ which requires using (possibly costly) bootstrap approaches \citep{FUH93}. Besides, the validity of such distributions remains usually asymptotic. Finally, even if this point has not been investigated, using an informative prior could maybe solve some identifiability problems \citep{ALL09}, when the dimension of $\bm \psi$ is high and/or data are poorly informative \citep{PUO09}.

A convenient prior for the transition matrix ${\bm{\psi }}$  can be obtained as the product of $r$ independent Dirichlet distributions, one for each row $\bm{\psi_i}$ of ${\bm{\psi }}$:
\begin{eqnarray}
\pi(\bm{\psi_i}) \propto \prod\limits_{j = 1}^{r} {\psi_{i,j}^{\gamma _{i,j}-1 } }.\label{Dirichlet.prior}
\end{eqnarray}
As the Dirichlet density is null outside the standard $(r-1)$-simplex, it is particularly suited as a prior distribution of probabilities vectors, that must fulfil conditions (\ref{transition.conditions}). Another well-known rationale for choosing a Dirichlet prior is that it can be seen as the reference posterior for a multinomial parameter given some virtual data of state-occupancy, whose sizes $\gamma_{i,j}-1$ can be interpreted as measures of the prior's strength \citep{MIN00}. However,  in absence of precise expert opinion in the remainder of this paper, uniform priors $(\gamma_{i,j}=1, \, \forall i,j$) were used for the examples shown hereinafter, following the recent recommendations from \citet{TUY09} based on symmetry requirements of posterior predictive distributions. 

\subsection{Complete sequence problem} \label{review.complete}
Transition probabilities estimation can easily be performed when complete states time-series (often alternatively called {\it panel data}) are available  for the $m$ individuals.
The estimation is based on the calculation, for every couple of states $(s_i, s_j)$, of the number of observed one-step transitions from state $s_i$ to state $s_j$:
\begin{eqnarray}
w_{i,j}   =  \sum\limits_{t = 1}^T {\sum\limits_{k = 1}^m {\mathbbm{1}_{ \left\{z_{(k,t - 1)} = s_i ,z_{(k,t)} = s_j \right\} } }}. \label{suff.statistics}
\end{eqnarray}
Full data likelihood can be written as a function of the sufficient statistics $w_{i,j}$ by observing that conditional on the row vector ${\bm{\psi_i}}=(\psi_{i,1}...\psi_{i,r})$, the vector ${\bm w_i}=(w_{i,1}...w_{i,r})$ is multinomial with parameters ${\bm{\psi _i}}$  and $\sum_{j = 1}^r {w_{i,j} }$. Therefore, the likelihood $L\left( {{\bm z}{\rm{|}}{\bm{\psi}} } \right)$  can be written as the product of $r$ multinomial terms:								
\begin{eqnarray}
L\left( {{\bm z}{\rm{|}}{\bm \psi} } \right)  =  \prod\limits_{i = 1}^r {\binom{\Sigma _{j}w_{i,j}}{w_{i,1} \ldots w_{i,r}} \ \psi _{i,1} ^{w_{i,1} }  \ldots \psi _{i,r} ^{w_{i,r} } } . \label{init.likelihood}
\end{eqnarray}
In a Bayesian framework, the estimation of transition probabilities given complete sequences is straightforward. The  inference problem consists in computing the posterior probability distribution of model parameters  $\pi ({\bm{\psi }}{\rm{|}}{\bm z})$  by updating the prior distribution $\pi ({\bm{\psi }})$ conditional to the observed data $\bm{z}$, through the Bayes formula:
\begin{eqnarray}
\pi({\bm{\psi }}{\rm{|}}{\bm z})  =  \frac{L\left( {{\bm z}{\rm{|}}{\bm \psi} } \right) \pi ({\bm{\psi }})}{\int_{\Omega} L\left( {{\bm z}{\rm{|}}{\bm \psi} } \right) \pi ({\bm{\psi }}) \ d{\bm \psi}}, \label{bayes.formula}
\end{eqnarray}
where $\Omega$ denotes the set of all possible values of  ${\bm \psi}$.
From (\ref{bayes.formula}), it can be seen that the prior~(\ref{Dirichlet.prior}) of $\bm{\psi }$ is conjugate, i.e. the posterior distributions of the ${\bm{\psi }}_i$'s are also Dirichlet distributions, with parameter vectors equal to ($\gamma _{i,1}  + w_{i,1} ,\ldots,\gamma _{i,r}  + w_{i,r}$). This is the well known Dirichlet-multinomial model.

\subsection{Incomplete sequence problem, ignorable DCM} \label{review.incomplete}

In the most general case of incomplete sequences problem, the estimation problem turns out to be more complicated.
Throughout this study, we will mostly consider the case where the Data Collection Mechanism (DCM) is ignorable, which means, in practice, that it can be neglected in the statistical data analysis. Besides simplicity purposes, this choice is essentially motivated by the framework and the background of our study, which is reliability analysis. Some elements about the more general cases of non-ignorable DCM will be provided in the next section.

Let $x_{(k,t)}$ be an auxiliary binary variable (missingness indicator) which is one if the observation is missing, zero if the state has been observed. The DCM is described by a complementary statistical model specifying  $\mathbbm{P}\left( x_{(k,t)} | \bm z, \bm z_{\texttt{mis}}, \bm{\eta } \right)$, i.e. the probability for an observation to be missing, depending on observed and unobserved data and (possibly) some other parameters $\bm{\eta }$.

Fulfilling two conditions is sufficient for ignorability \citep{GEL04}: the first one states the independence between the parameters of the DCM and the main model (here $\bm{\eta }$ and $\bm{\psi }$ respectively), the second one asserts that the probability that an observation is missing does not depend on missing data (MAR: \textit{missing at random} condition). The first condition is generally easily checked, while the second one highly depends on the context of the statistical study. For instance, in capture-recapture experiments the probability of recapture may depend or not on the state of the individual (e.g. younger animals can be more easily captured than older ones). In longitudinal medical surveys the health state of a patient can prevent him from going to a periodical visit (e.g. in case he/she is hospitalized).
In an industrial reliability framework, and in particular in the specific context of EDF (Electricit\'{e} de France), the presence of missing data is mainly due to the impossibility of simultaneously surveying the whole population of components for cost or system availability reasons. This motivates our choice to mainly focus on ignorable DCM situations.

Let us now come back to our estimation problem. In incomplete sequences problems, the likelihood has a highly complex expression. It is the product of $m$ terms which are the probabilities to observe each one of the $m$ sequences. Whilst writing the term related to an incomplete sequence, one must consider all possible values of the unknown observations. For example, the probability of the sequence $(s_1 ,s_1 ,\bullet,\bullet,s_3)$ must be written by taking into account all possible three-steps paths from state $s_1$ to state $s_3$:
\begin{eqnarray*}
\mathbbm{P}(s_1 ,s_1 ,\bullet,\bullet,s_3) \propto \sum\limits_{i=1}^{r}\left[ \psi _{1,i}\sum\limits_{j=1}^{r}\psi _{i,j}\psi
_{j,3}\right].
\end{eqnarray*}
Estimation methods dealing directly with the likelihood expression may be quite tricky to perform \citep{DEL99}. On the other hand, Bayesian inference can elegantly be performed by means of a Gibbs sampler.

This procedure is particularly adapted to the cases where the posterior distribution of model parameters would be more easily determined if data were fully observed. Missing data are considered as additional model parameters $z_{\texttt{mis}(k,t)}$ and, within the Gibbs sampling, an additional step is performed to simulate them, thus completing the data set. This technique is usually known as \textit{data augmentation} \citep{ROB04}. Note that Gibbs sampling may be viewed as the Bayesian mirror of Stochastic Expectation-Maximization (SEM) algorithms based on a similar mechanism \citep{DEL99}.

In our case the augmented data set, say ${\bm y}$, is the set of the completed state sequences for all individuals:
\begin{eqnarray*}
y_{(k,t)}  =  z_{(k,t)} ~ \text{if $z_{(k,t)}$ is observed},
\end{eqnarray*}
and
\begin{eqnarray*}
y_{(k,t)}  =  z_{\texttt{mis}(k,t)} ~ \text{otherwise.}
\end{eqnarray*}
The Gibbs sampler algorithm for the incomplete sequence problem can be viewed as a particular case of the more general method for the Arnason-Schwarz capture-recapture model \citep{MAR07}.
We first initialize the algorithm by arbitrarily completing state sequences. Then at each step $h=1, 2,\ldots$, we perform the following two-step procedure:
\begin{enumerate}
\item drawing new parameter values, conditional on the augmented data ${\bm y}^{[h-1]}$:
\begin{eqnarray*}
{\bm{\psi }}_i^{[h]} {\rm{|}}{\bm y}^{[h-1]}  \sim  \Dir\left(\gamma _{i,1}  + w_{i,1}^{[h-1]} ,...,\gamma _{i,r}  + w_{i,r}^{[h-1]}\right), 
\end{eqnarray*}
where $w_{i,j}^{[h-1]}$ are the sufficient statistics (\ref{suff.statistics}) evaluated from current completed sequences ${\bm y}^{[h-1]}$;
\item drawing missing data  $z^{[h]}_{\texttt{mis}(k,t)}$ conditional to the current values ${\bm{\psi }}^{[h]}$  of model's parameters (data augmentation step). This can be done by sampling from a conditional categorical distribution defined by the following probabilities:
\begin{eqnarray*}
\mathbbm{P} \left( {y^{[h]}_{(k,1)} = s_j {\rm{|}}y^{[h - 1]}_{(k,2)}=s_i,{\bm{\psi }}^{[h]} } \right) & \propto &  \psi _{j,i}^{[h]},  \text{for $t=1$},
\end{eqnarray*}
\begin{eqnarray*}
\mathbbm{P} \left( {y^{[h]}_{(k,T)} = s_j {\rm{|}}y^{[h]}_{(k,T - 1)}=s_i,{\bm{\psi }}^{[h]} } \right) & \propto & \psi _{i,j}^{[h]},  \text{for $t=T$},
\end{eqnarray*}
and
\begin{eqnarray*}
\mathbbm{P} \left( {y^{[h]}_{(k,t)} = s_j {\rm{|}}y^{[h]}_{(k,t - 1)}=s_{i_1},y^{[h - 1]}_{(k,t + 1)}=s_{i_2},{\bm{\psi }}^{[h]} } \right) & \propto & \psi _{i_1,j}^{[h]}  \cdot \psi _{j,i_2}^{[h]},  \text{otherwise}.
\end{eqnarray*}
\end{enumerate}
The computational method shown above is quite general and easy to implement. On the other hand, the more incomplete the sequences are, the more additional parameters are required and the more the data augmentation step becomes time-consuming. This issue will be illustrated later on in the example of Section \ref{example1}. A technique to accelerate this step, consisting in simulating blocks of consecutive missing data instead of one datum at a time, is proposed by \citet{DUP07}.

A particularly interesting case of incomplete sequence problem occurs when each individual is observed just once over the observation period. This can happen in industrial reliability when the data come from the first survey of operating machines, as in the real-world example of Section \ref{complete.exemple}, or from destructive controls \citep{PAS04}. Then let $t_k$ (with $1<t_k<T$) be the time when the individual $k$ has been observed and $s_j$ be the observed state. The state sequences takes the form:
\begin{eqnarray*}
\bullet,\ldots,\bullet,s_j ,\bullet,\ldots,\bullet.
\end{eqnarray*}
In that case, it can be shown (proof in \ref{app:proof}) that the likelihood $L\left( {{\bm z}{\rm{|}}{\bm \psi}} \right) $ has the general expression:
\begin{eqnarray}
L\left( {{\bm z}{\rm{|}}{\bm \psi}} \right) \propto \prod\limits_{t=1}^{T}\prod\limits_{j=1}^{r}p_{j}(t)^{n'_{j}(t)}. \label{likelihood.once.observed}
\end{eqnarray}
In the formula above, $p_{j}(t)$ is the unconditional probability for the system to be in state $s_j$ at time $t$ and $n'_{j}(t)= \sum_{i=1}^{m} \mathbbm{1}_{\{z_{(i,t)}=s_j\}}$ is the number of times the state $s_j$ has been observed at time $t$ in the data sample $\bm{z}$.
It has to be noticed that the expression of the likelihood depends on sufficient statistics $n'_{j}(t)$ and the statistical problem is equivalent to the aggregate data problem considered hereinafter.
In this particular case, Bayesian estimation can be performed using the Gibbs sampler  described above or the Metropolis-Hastings procedure we carry out for the aggregate data problem in Section \ref{review.aggregate}.

\subsection{Incomplete sequence problem, non-ignorable DCM} \label{review.incompleteMNAR}
Let us now consider the more general case where DCM is non ignorable. 

This problem has been studied in detail \citep[Chapters 6-10]{LIT87} in particular within the framework of longitudinal medical surveys: indeed, for different reasons, patients can leave the study permanently (dropout) or temporarily (intermittent missing).
Using the same notation as in the previous subsection, let $\bm{y}_k$ be a complete data sequence for the individual $k$ (while ${\bm z}_k$ denotes the actually observed sequence). The different ways for coping with MNAR (missing not at random) observations rely, from a technical point of view, on the way the \textit{full-data} likelihood $L( {\bm y}_k, {\bm x}_k| {\bm \psi }, {\bm\eta})$ is factorized. Three types of factorization are usually proposed:
\begin{eqnarray*}
L({\bm y}_k|{\bm x}_k,{\bm \psi }) \cdot L({\bm x}_k| {\bm \eta }) ~~\text{(pattern mixture model),}
\end{eqnarray*}
\begin{eqnarray*}
L({\bm y}_k|{\bm \psi }) \cdot L({\bm x}_k| {\bm y}_k , {\bm\eta}) ~~\text{(selection model),}
\end{eqnarray*}
and
\begin{eqnarray*}
\int L({\bm y}_k | {\bm x}_k,{\bm v}_k,{\bm \psi }) \cdot L({\bm x}_k | {\bm v_k}, {\bm \eta })\cdot f({\bm v}_k |{\bm\lambda}) \,d{\bm v_k}~~\text{(shared parameter model).}
\end{eqnarray*}

The formulations above can be complexified, by considering the influence of covariates in both the main and the missingness models.

In the pattern mixture framework \citep{LIT93}, the analyst models the conditional distribution of the observable outcome, given its observation pattern, and the distribution of the different patterns. As a matter of fact, the data are stratified (each pattern determines a stratum) and the main parameters ${\bm \psi}$ are estimated in each stratum.

The selection factorization, first introduced by \citet{RUB76}, instead, focuses on the dependence between the missingness and the actual value of the observable variable (in our case the state of the individual). This scheme explicitly copes with the distribution of the complete data ${\bm y}$ conditional on the main parameter of the model, here ${\bm \psi}$. The DCM parameters ${\bm \eta }$ are easy to interpret and provide additional valuable information to the analyst.

In the shared-parameter scheme \citep{WU88}, the missing mechanism is indirectly related to the observable variable through a latent variable ${\bm v}$, depending on some additional parameters ${\bm\lambda}$.

In the particular framework of the estimation of transition probabilities, \citet{COL05} considered categorical quality-of-life data in cancer clinical trials, using a selection factorization. Transition probabilities $\psi_{i,j}$ and missingness probabilities $\eta_i=\mathbbm{P}(x_{k,t}=1|y_{(k,t)}=s_i)$ both depend on observable covariates.

The Arnason-Schwarz model \citep{DUP95, MAR07}, also based on a selection factorization, provides an elegant Bayesian solution in the case where the $\eta_i$'s do not depend on covariates. In this case, a natural choice of the prior for each one of the $\eta_i$'s is a Beta pdf: $\Be(\alpha_i,\beta_i)$. The Gibbs algorithm for estimating the posterior of $({\bm\eta},{\bm\psi})$ is straightforward as, conditional on the the complete data $\bm y$, both posterior distributions of ${\bm\eta}$ and ${\bm\psi}$ are explicit. The detailed description of the two steps of the algorithm (data augmentation and parameters estimation) is given in \ref{app:GibbsMNAR}.

\subsection{Aggregate data problem} \label{review.aggregate}
In many real-life problems, we do not follow individuals passing from state to state and the only available data for estimating transition probabilities are aggregate data ${\bm n}$, i.e. the number of individuals $n_i(t)$ being in a given state $s_i$ at a given time $t$. Any track of individual trajectories is lost. That may occur in practice when a population of $m$ individuals has been followed over an observation period but the original aim of the survey was simply having the fractions of the population in particular states. State sequences have thus been considered as raw data and discarded. Examples in sociology and population dynamics were highlighted by \citet{BAR73} and \citet{POL73}, among others. Applications in credit rating were recently studied by \citet{JON05}.

The inference problem has been formalized by \citet{LEE68}. Conditional on the probability vector ${\bm p}(t) = {\bm p}(0) \cdot {\bm{\psi }}^t$, the data vector ${\bm n}(t) = (n_1 (t),n_2 (t),...,n_r (t){\rm{)}}$ is multinomial with parameters ${\bm p}(t)$  and $\sum_{j = 1}^r {n_j (t)}$.
The likelihood $ L\left( {{\bm n}{\rm{|}}{\bm{\psi }}} \right)$ can then be written as the product of $T$ independent terms:
\begin{eqnarray}
L\left( {{\bm n}{\rm{|}}{\bm{\psi }}} \right)  =  \prod\limits_{t = 1}^T {\frac{{\left( {\sum\nolimits_j {n_j (t)} } \right){\rm{ }}!}}{{\prod\limits_{j = 1}^r {n_j (t)!} }}\prod\limits_{j = 1}^r {p_j (t)^{n_j (t)} } } \label{likelihood}.
\end{eqnarray}
\citet{LEE68} focused on obtaining point estimates of the matrix $\bm{\psi}$ and in particular the posterior mode of $\pi ({\bm{\psi }}{\rm{|}} \bm{n})$ by maximizing the product of the likelihood (\ref{likelihood}) and $r$ independent Dirichlet priors (\ref{Dirichlet.prior}), one for each row of $\bm{\psi}$.

In the same frequentist context, \citet{MAC77} then \citet{KAL84} were among the main authors who developed generalized least square estimators to remedy the difficulty of the maximum likelihood estimation, because of the untractability of $L( {{\bm n}{\rm{|}}{\bm{\psi }}})$. Under mild conditions on the stationary matrix $\bm\psi$, \citet{KAL84} obtained general consistency results and asymptotic $r(r-1)-$variate normality (in $T$ and $N=\sum_{j=1}^r n_j(t)$) for the estimated vector $\bm{\psi}_{\text{row}}$ of entries in $\bm\psi$ written rowwise, i.e. $\bm{\psi}_{\text{row}}=(\psi_{1,1},\ldots,\psi_{1,r-1},\psi_{2,1},\ldots,\psi_{r,r-1})$. \citet{LAW84} gave conditions on functions of interest for which the information loss due to aggregation is asymptotically negligible with respect to the estimation based on complete sequences. In a specific reliability framework, \citet{GOU11} recently provided a methodology to estimate such functions of interest (e.g. survival probability, sojourn time in a state) .

In a Bayesian context, the inference problem can be solved by using a Metropolis-Hastings (MH) algorithm to construct a sample  of matrices of $\Omega: {\bm{\psi }}^{[0]} ,{\bm{\psi }}^{[1]} ,\ldots,{\bm{\psi }}^{[h]} ,\ldots$, asymptotically drawn from the posterior $\pi ({\bm{\psi}}{|\bm n})$, by sampling at each step $h$ a candidate vector ${\bm{\psi }}^{[h]*}$ from a given distribution function $J(\cdot{\rm{|}}{\bm{\psi }}^{[h - 1]} )$. The candidate is accepted with probability:
\begin{eqnarray}
\rho({\bm{\psi}} ^{[h]*} {\rm{|}} {\bm{\psi}} ^{[h - 1]} ) = 1 \wedge \frac{{\pi ({\bm{\psi }}^{[h]*}{|\bm n} )}}{{\pi ({\bm{\psi }}^{[h - 1]} {|\bm n})}} \cdot \frac{{J({\bm{\psi }}^{[h - 1]} {\rm{|}}{\bm{\psi }}^{[h]*} )}}{{J({\bm{\psi }}^{[h]*} {\rm{|}}{\bm{\psi }}^{[h - 1]}) }},
\label{acceptance.probability}
\end{eqnarray}
i.e. the acceptance of the candidate is the result of a Bernoulli trial of probability $\rho({\bm{\psi}} ^{[h]*} {\rm{|}} {\bm{\psi}} ^{[h - 1]} )$.\\
The {\it instrumental} density function $J(\cdot {\rm{|}}{\bf{\bm\psi }}^{[h - 1]} )$  allows a random exploration of the space of parameters. The convergence of the chain to the target distribution is proved for any arbitrary function $J(\cdot {\rm{|}} \cdot)$ which satisfies mild regularity conditions \citep{ROB04}. In the present case, a comfortable instrumental function is the product of $r$ independent Dirichlet distributions 
$\text{Dir}(d_i  \cdot {\bm{\psi }}_{i}^{[h - 1]})$, where $d_i$ is a positive (scalar) constant. This is a usual case of {\it controlled} MCMC \citep{AND08}. As the Dirichlet density is null outside the standard $(r-1)$-simplex, all candidates drawn by the instrumental functions automatically respect constraints (\ref{transition.conditions}). Obviously the support of $J$ contains the support of the posterior distribution and the chain is a reversible jump MCMC. 

It can easily be seen that the mean of each 
of the $r$ Dirichlet instrumental densities is ${\bm{\psi }}_i^{[h - 1]}$, i.e. the candidate matrix is sampled from a probability function which is ``centered'' on the last retained matrix. The variance terms of the covariance matrix, equal to $\psi _{i,j}^{[h - 1]}  (1 - \psi _{i,j}^{[h - 1]})  / (d_i+1)$, depend on the shape parameters $d_i$ which can be interpreted as tuning coefficients that rule the distance of exploration from the current state of the MCMC chain to the next proposed one.

We notice that, as the expressions of the likelihoods (\ref{likelihood.once.observed}) and (\ref{likelihood}) are formally the same, up to a proportionality constant, the MH procedure described above can also be used in the interesting case of incomplete sequences when each individual has been observed only once. Such examples are treated in the next paragraphs. 
\subsection{Real industrial case-study: survey of power station turbines}\label{complete.exemple}

In the example shown hereby, a discrete Markov model has been used to describe the propagation of transverse cracks on steam turbine shafts. This phenomenon has been first observed on EDF facilities in late 90's and since then periodical non-destructive controls are made to measure crack depths. For a description of the technical problem and available survey data, see \citep{GAR06}.
The most important identified explanatory variable is the time spent by the turbine in hot shutdown condition. For the purpose of our study, the time has been discretized in equally long intervals. 
Cracks depths are classified in four states $s_1 \ldots s_4$ associated to growing crack lengths. The modelling of cracks growth by discrete Markov schemes is quite common, e.g. \citep{ROH00}.

We assume that the process is irreversible, which is physically correct as crack lengths cannot decrease. Thus, the transition matrix is upper-triangular and consequently, $\psi_{4,4}=1$. \\
We made the hypothesis that all turbines are in state $s_1$ when putting-into-service at the beginning of the study. Manufacture and acceptance controls justify this hypothesis.
A set of data collected between 1998 and 2001 has been analyzed. The data come from 68 turbines from 24 EDF power plants. Each turbine is observed only once for a given value of $t$ between 2 and 7. Given the uniformity of EDF French generation facilities (same design, operating conditions and maintenance policy for all units), we can assume that observed data are i.i.d.

The results of MCMC estimation, using the Gibbs sampler described in Section~\ref{review.incomplete} (second half run of 10000 iterations), are shown in Table 1 (left). The application of the MH algorithm described above leads to the same results.

The data set has been enriched between 2001 and 2004 with new crack measures ($t$ between 2 and 7). 38 turbines among the 68 previously observed were inspected for the second time and two for the first time. Some of the collected data are redundant: this happens when for the first and the second observation the corresponding times spent in hot shutdown condition fall into the same interval. Finally, 17 new exploitable observations can be added to the data set. The estimation of transition probabilities gives the results shown in Table 1 (right).
\begin{table}[h!]
\label{reviewtable1}
\begin{center}
\begin{tabular}{ccc}
& Data set 1 & Data set 2 \\ 
\hline
\begin{tabular}{c}
\\ 
$\psi _{1,1}$ \\ 
$\psi _{1,2}$ \\ 
$\psi _{1,3}$ \\ 
$\psi _{1,4}$ \\ 
$\psi _{2,2}$ \\ 
$\psi _{2,3}$ \\ 
$\psi _{2,4}$ \\ 
$\psi _{3,3}$ \\ 
$\psi _{3,4}$%
\end{tabular}
& 
\begin{tabular}{ccc}
Mean & St.\ Dev. & 95\%\ CI\\
\hline 
0.637 & 0.042 & [0.551,\, 0.719]\\ 
0.306 & 0.050 & [0.208,\, 0.405]\\ 
0.044 & 0.034 & [0.002,\, 0.127]\\ 
0.012 & 0.011 & [0.000,\, 0.041]\\ 
0.713 & 0.088 & [0.538,\, 0.884]\\ 
0.250 & 0.087 & [0.079,\, 0.418]\\ 
0.037 & 0.032 & [0.001,\, 0.119]\\ 
0.872 & 0.097 & [0.627,\, 0.995]\\ 
0.128 & 0.097 & [0.005,\, 0.373]
\end{tabular}
& 
\begin{tabular}{ccc}
Mean & St.\ Dev. & 95\%\ CI\\ 
\hline
0.655 & 0.040 & [0.573,\, 0.728]\\ 
0.278 & 0.049 & [0.185,\, 0.374]\\ 
0.056 & 0.036 & [0.003,\, 0.133]\\ 
0.012 & 0.011 & [0.000,\, 0.041]\\ 
0.774 & 0.075 & [0.636,\, 0.921]\\
0.197 & 0.075 & [0.054,\, 0.341]\\ 
0.029 & 0.024 & [0.001,\, 0.087]\\
0.910 & 0.071 & [0.730,\, 0.996]\\ 
0.090 & 0.071 & [0.004,\, 0.270]
\end{tabular} \\
\hline
\end{tabular}
\caption{Turbine cracks example. MCMC estimations of transition matrix $\bm\psi$ using the first data set (left, individuals observed only one time) and the second data set (right). Here, the bounds of the posterior 95\% credibility intervals (CI) are the quantiles of probabilities 0.025 and 0.975 respectively.}

\end{center} 
\end{table}

We can notice that in this case the posterior variance has been very lightly reduced by incorporating the information conveyed by the new data. Given the posterior samples of transition probabilities, some quantities of practical interest in industrial reliability have been sampled: the unconditional probabilities of the four states, as a function of time, and the expected number of steps before the system reaches the absorbing state $s_4$. As $s_4$ can be interpreted as a ``failure'' condition, the expected time to absorption is here the classical MTTF (Mean Time To Failure). Notice that here the term ``failure'' just means that the crack has reached a given length, arbitrarily chosen for the purposes of this paper.

The calculation of state probabilities using Equation (\ref{probabilities.0}) is straightforward.
To evaluate the MTTF we made use of a well known property of absorbing Markov chains \citep[chap. 11]{GRI97}. If we consider the matrices:
\begin{eqnarray*}
{\bm{\zeta}} & = & \left( {\begin{array}{*{20}c}
   {\psi _{1,1} } & {\psi _{1,2} } & {\psi _{1,3} }  \\
   0 & {\psi _{2,2} } & {\psi _{2,3} }  \\
   0 & 0 & {\psi _{3,3} }  \\
\end{array}} \right) \ \ \text{and} \ \ 
{\bm{I}} \ = \ \left( {\begin{array}{*{20}c}
   1 & 0 & 0  \\
   0 & 1 & 0  \\
   0 & 0 & 1  \\
\end{array}} \right),
\end{eqnarray*}
the matrix ${\bm{I}}-{\bm{\zeta}}$  has an inverse and each component $t_i^*$ of the row vector
\begin{eqnarray*}
\bm{t^*} & = & (1,1,1) \cdot \left( {\bm{I}}-{\bm{\zeta}} \right)^{ - 1} 
\end{eqnarray*}
is the expected number of steps before absorption, given that the initial state was $s_i$. In our case the MTTF is then the first component of the vector $t^*$.\\
\begin{figure}
\centering
\includegraphics[width=13.5cm,height=8cm]{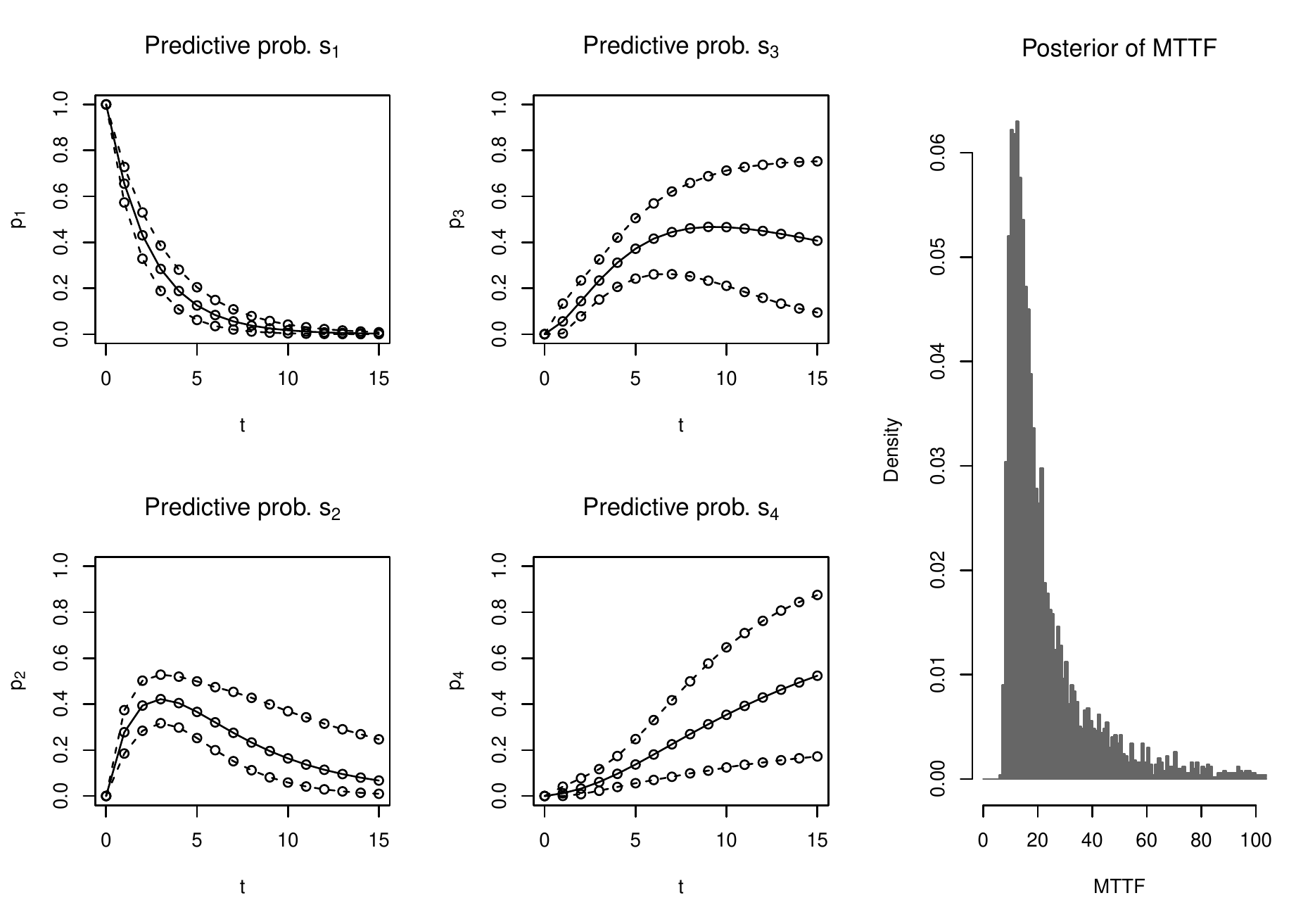}
\caption{Turbine cracks example. Predictive 95\% credibility intervals of state probabilities (left) and predictive distribution of the MTTF (right).}
\label{review.example.figure}
\end{figure}
Figure \ref{review.example.figure} shows the 95\% credibility intervals of the predictive state probabilities for discretized time $t$ extended up to 15 and the histogram of 5000 samples from the predictive distribution of MTTF. Concerning state probabilities, we can notice that $p_1$ credibility intervals are narrower than other state probabilities as, according to our hypotheses of an irreversible process and initial state $s_1$, $p_1 (t) = \psi _{1,1}^t$
which mean that the uncertainty over $p_1$ only depends on uncertainty over $\psi _{1,1}^{}$  (and no other transition probability). 
The long tail in the MTTF distribution (which is even longer than shown in the figure) is due to the high values (close to 1) of the posterior distribution of $\psi _{3,3}$.

\textbf{Remark}. We stress that, even if the data come from real surveys, the study shown hereinbefore is given for exemplary purposes only and neither results nor methodology must be extrapolated to make any general conclusion about EDF risk assessment policies.

\subsection{A four-dimensional simulated case study}\label{example1}

The purpose of the previous example was to give a practical use-case of application of the estimation algorithms in a poorly informative data context. The next example will deal with more general computational issues. In particular, we will compare the performances of Gibbs and MH algorithms in the case where individuals are observed only once.
Following a case-study from \citet{LEE68}, we consider the following transition matrix:
\begin{eqnarray}\label{psi.lee}
{\bm{\psi }}_{o} & = & \left( {\begin{array}{*{20}c}
   {0.6} & {0.4} & 0 & 0  \\
   {0.1} & {0.5} & {0.4} & 0  \\
   0 & {0.1} & {0.7} & {0.2}  \\
   0 & 0 & {0.1} & {0.9}  \\
\end{array}} \right).
\end{eqnarray}

\begin{figure}[hbtp]
\centering
\includegraphics[height=8cm,width=10cm]{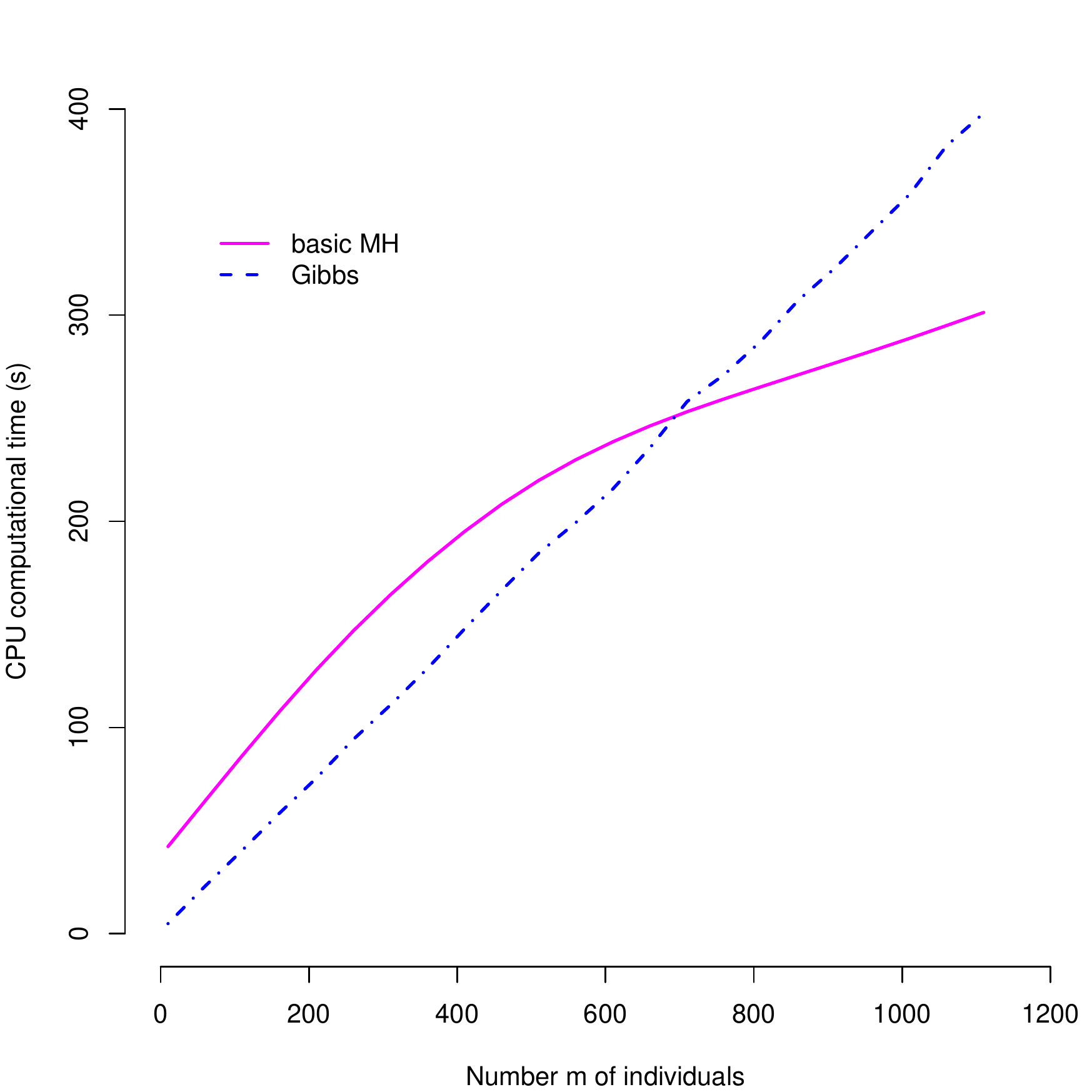}
\caption{Case study of \citet{LEE68}. Mean CPU time needed to reach quasi-stationarity (in the sense of the Brooks-Gelman rule of thumb) as a function of the number $m$ of individuals (one individual being associated to a single true observation). Data have been generated according to the four-state transition matrix~(\ref{psi.lee}).}
\label{comparison-gibbs-mh-1}
\end{figure}

First, complete state sequences for $m\in\{10,\ldots,1200\}$ individuals have been generated for $T=20$ observation periods, under the hypothesis that at $t=0$ the initial vector probability is $(3/4, 1/4, 0, 0)$. 
Then, given complete sequences, a single observation per individual has been randomly selected, thus obtaining  incomplete sequences.
Finally, for each $m$ we used the Gibbs and the Metropolis-Hastings algorithms described above. 
The convergence has been checked using the Brooks-Gelman statistic \citep{BRO98} computed on three parallel chains  and a visual inspection of the chains. A classic rule of thumb (RT) is to suppose quasi-stationarity once the statistic stably remains under 1.1 \citep{BRO98}. The precision in estimation was measured using the relative absolute error matrix between the elements of ${\bm{\psi }}_o$ and a progressive Monte Carlo posterior estimate of $\bm\psi$. In each case, it has been obtained by using the second half run of Metropolis-Hastings iterations and Gibbs iterations after the burn-in periods determined by Brooks-Gelman RT respectively. Parameters $d_i$ were sampled uniformly in [100,2500].
For a same estimation error of at most 5\% per element, the CPU time observed on a 2.8 GHz CPU (Xeon) machine before the RT is fulfilled has been plotted in Figure \ref{comparison-gibbs-mh-1} as a function of $m$. Plots are smoothed over 30 repetitions of the  algorithms. Clearly, the increasing number of missing data makes Gibbs less competitive than MH: after $m=700$, conditional sampling of individuals requires more CPU time than our basic MH. The number of missing data to be simulated increases linearly with the total number $m$ of individuals, as individuals could be observed only once throughout their lifespan. This explains the linear behavior of the Gibbs CPU time.

The efforts of the practitioner should then concentrate on improving the mixing of Gibbs and MH algorithms to diminish their burn-in period. The development of acceleration methods has been the subject of a large number of works, reviewed in \citep{GILKS96,MIRA2003,GENTLE04}. Techniques such as blocking \citep{ROBERTS97}, which consists in updating multivariate blocks of (often highly correlated) parameters, were shown to be efficient to accelerate Gibbs algorithms in conjugate models \citep{ISCH01,ACC09}, although their implementation often remains case-specific \citep{SAR00} and can sometimes slow the sampler's convergence \citep{ROBERTS97}. Alternatively, the multi-move Gibbs sampler \citep{CARTER94}, which was developed for Markov switching state-space models, proved to be more efficient than the single-move Gibbs sampling, although its filtering aspects might be time-consuming \citep[chap. 11, pp. 342-344]{FRUWIRTH06}. More recently, cheaper approximations of the Gibbs sampler using best linear predictors have been carried out \citep{NOTT2005}.

In the specific case of MH algorithms, the slow mixing and the computational cost are less due to the dimension of the problem than the difficulty of eliciting instrumental distributions to efficiently explore the parameter space. The remainder of this article is specifically dedicated to an empirical exploration of adaptive approaches aiming to improve this feature.


\section{Accelerating the MH algorithm using adaptive approaches}\label{research}

Heuristically, implementing an adaptive MCMC consists in sequentially tuning the transition kernel using the knowledge of past iterations, in an automated way during the simulation, in order to improve the mixing rate \citep{AND08}. In the particular case of our class of MH algorithms, this means modifying the product of Dirichlet densities chosen as the instrumental distribution $J$ for the  MH algorithm introduced in Section~\ref{review.aggregate}. Each successive instrumental distribution is ideally selected  such that parallel sampling can explore a large part of the parameter space, especially in the first steps of the algorithm.

Recently, a rich literature has been dedicated to these approaches, and is especially focused on the preservation of the ergodicity of the adaptive chains towards the stationary distribution, which is not automatically ensured by automated tunings. Seminal works on this subject are due to Roberts and Rosenthal (\citeyear{ROBS07,ROBS09}) and Andrieu et al. \citep{AND06,AND07,AND08}. 
These theoretical works also led to interesting software developments \citep{ROS07,VIH10}.

Assuming $\Gamma_i$ are indices chosen in some collection ${\cal{Y}}$ based on past algorithm output, we denote by  $K_{\Gamma_i}$ the transition kernel updating $\bm\psi^{[i]}$ to $\bm\psi^{[i+1]}$:
\begin{eqnarray}
K_{\Gamma_i}\left(\bm\psi,\bm\psi'\right) & = & \rho_{\Gamma_i}\left(\bm\psi,\bm\psi'\right) J_{\Gamma_i}\left(\bm\psi'|\bm\psi\right) + \int \left(1-\rho_{\Gamma_i}\left(\bm\psi,\bm\epsilon\right)\right) J_{\Gamma_i}\left(\bm\epsilon|\bm\psi\right) d\bm\epsilon \ \delta_{\bm\psi}\left(\bm\psi'\right),
\end{eqnarray}
where $\delta_{\bm\psi}$ is the Dirac measure in $\bm\psi$ and 
\begin{eqnarray*}
\rho_{\Gamma_i}\left(\bm\psi,\bm\psi'\right) & = & 1 \wedge \frac{\pi(\bm\psi'|{\bm z})J_{\Gamma_i}\left(\bm\psi|\bm\psi'\right)}{\pi(\bm\psi|{\bm z})J_{\Gamma_i}\left(\bm\psi'|\bm\psi\right)}.
\end{eqnarray*}
Basically, the ergodicity and stationarity properties of an adaptive MH algorithm can be ensured if the amount of adapting progressively {\it diminishes}, in the sense that the kernel parameters are modified by smaller and smaller quantities, or if the probability of adaptation $\rho_{\Gamma_i}$ decreases towards zero as $i\to\infty$ \citep[Theorem 5]{ROBS07}. In the framework considered here, such adaptations could be based on eliciting  vanishing adaptations for the parameters $(d_i)_{1\leq i\leq r}$. These approaches would however be limitative since each $d_i$ characterizes the marginal distribution of row $i$, hence they do not explore the correlations between the rows. Therefore, the approach proposed here focuses on this particular aspect.

In the following, assuming we are at step $h>1$ of the MH algorithm, we propose two ways of building an adaptive instrumental distribution $\bm\psi^{[h]\ast}\sim J_{h}$ (denoting $J_{\Gamma_h}= J_h$ in the following for simplicity) taking advantage of a $\sigma-$algebra ${\cal{F}}_{h-1}$ generated by the succession of sampled parameter matrices $\bm\psi^{[0]},\ldots,\bm\psi^{[h-1]}$. Both using a (small) fixed number $p$ of basic MH iterations, these approaches explore correlations between the rows in the instrumental sampling.  

In our first approach (DCS-MH), we attempt to summarize the correlations within $(\bm\psi_1,\ldots,\bm\psi_r)$ by simply capturing the correlations between the diagonal elements of $\bm\psi$.

In our second method (RCS-MH), we generalize the first method  replacing the $r-$vector of diagonal elements by $r$ elements whose position is randomly sampled within each vector $\bm\psi_i$. Doing so, we hope to capture more efficiently the dependency between the $\bm\psi_i$ and accelerate the DCS-MH algorithm.

\paragraph{Diagonal correlated sampling (DCS-MH)} 
\rule[0.5ex]{6cm}{0.1mm}
\vspace{0.1cm}
\small

\texttt{At iteration $h\gg p$ (large enough):} 
\texttt{
\begin{enumerate}
\item denote $ \{\tilde{\bm\psi}^{[1]}, \dots, \tilde{\bm\psi}^{[p]}\}$ the set of last $p$ non-identical sampled matrices in 
($\bm\psi^{[0]},\ldots,\bm\psi^{[h-1]}$);
\item for $i=1,\ldots,r$
\begin{description}
\item[(i)] denote $\tilde{\bm\psi}_{i,i} = (\tilde{\psi}^{[1]}_{i,i},\ldots,\tilde{\psi}^{[p]}_{i,i})$ the $p-$vector of replicates of the $i-th$-diagonal element;
\item[(ii)] compute ${\bf u_i}=\hat{F}_{i}(\tilde{\bm\psi}_{i,i})$ where $\hat{F}_{i}$ is the empirical marginal cdf of $\tilde{\bm\psi}_{i,i}$
\end{description}
\item estimate the Pearson correlation $\bf R^{[h]}$ of $(\bf u_1,\ldots,\bf u_r)$;
 \item sample a candidate vector
$\bm\psi^{[h]\ast}_{\text{diag}}$ of diagonal elements
$\psi_{1,1}^{[h]\ast}, \dots, \psi_{r,r}^{[h]\ast}$ using:
\begin{description}
\item[(i)] a Gaussian copula, the parameter of which is $\bf R^{[h]}$,
\item[(ii)] Beta marginal distributions
${\displaystyle \text{Be}\left(d_i\cdot\psi_{i,\,i}^{[h-1]},\,
d_i\left(1-\psi_{i,\,i}^{[h-1]}\right)\right)}$;
\end{description}
\item for $i=1,\ldots,r$ 
\begin{description}
\item[(i)] sample $\psi^{[h]\ast}_{i,1},\ldots,\psi^{[h]\ast}_{i,i-1},\psi^{[h]\ast}_{i,i+1},\ldots,\psi^{[h]\ast}_{i,r}$ from:  
\begin{eqnarray*}
 \Dir\left(\frac{\psi_{i,\,1}^{[h-1]}}{1-\psi_{i,\,i}^{[h-1]}}\cdot d_i,\dots,\frac{\psi_{i,\, i-1}^{[h-1]}}{1-\psi_{i,\,i}^{[h-1]}}\cdot d_i,\frac{\psi_{i,\,i+1}^{[h-1]}}{1-\psi_{i,\,i}^{[h-1]}}\cdot d_i,\dots,\frac{\psi_{i,\,r}^{[h-1]}}{1-\psi_{i,\,i}^{[h-1]}}\cdot d_i\right);
\end{eqnarray*}
\item[(ii)] for $j\neq i$, renormalize each $\psi_{i,j}^{[h]\ast}$ by multiplying  with $1-\psi_{i,\,i}^{[h]\ast}$. \\
\end{description}
\end{enumerate}
\rule[0.5ex]{\textwidth}{0.1mm}
}

\normalsize

\paragraph{Randomized correlated sampling (RCS-MH)} 
\rule[0.5ex]{5.5cm}{0.1mm}
\vspace{0.1cm}
\small 

\texttt{At iteration $h\gg p$ (large enough):} 
\texttt{
\begin{enumerate}
\item same as step 1 in DCS-MH;
\item sample (with replacement) a $r-$vector $I\in\{1,\ldots,r\}$ of random indicators;
\item for $i=1,\ldots,r$
\begin{description}
\item[(i)] denote $\tilde{\bm\psi}_{i,I_i} = (\tilde{\psi}^{[1]}_{i,I_i},\ldots,\tilde{\psi}^{[p]}_{i,I_i})$ the $p-$vector of replicates of the $(i,I_i)-th$ matrix element;
\item[(ii)] compute ${\bf u_{i}}=\hat{F}_{i}(\tilde{\bm\psi}_{i,I_i})$ where $\hat{F}_{i}$ is the empirical marginal cdf of $\tilde{\bm\psi}_{i,I_i}$
\end{description}
\item same as step 3 in DCS-MH; 
\item sample a candidate vector
$\bm\psi^{[h]\ast}_{\text{rand}}$ of elements
$\psi_{1,I_1}^{[h]\ast}, \dots, \psi_{r,I_r}^{[h]\ast}$ following the same main idea as in DCS-MH method
\item For $i=1,\ldots,r$ 
\begin{description}
\item[(i)] sample $\psi^{[h]\ast}_{i,1},\ldots,\psi^{[h]\ast}_{i,I_i-1},\psi^{[h]\ast}_{i,I_i+1},\ldots,\psi^{[h]\ast}_{i,r}$ from:  
\begin{eqnarray*}
 \Dir\left(\frac{\psi_{i,\,1}^{[h-1]}}{1-\psi_{i,\,I_i}^{[h-1]}}\cdot d_i,\dots,\frac{\psi_{i,\, I_i-1}^{[h-1]}}{1-\psi_{i,\,I_i}^{[h-1]}}\cdot d_i,\frac{\psi_{i,\,I_i+1}^{[h-1]}}{1-\psi_{i,\,I_i}^{[h-1]}}\cdot d_i,\dots,\frac{\psi_{i,\,r}^{[h-1]}}{1-\psi_{i,\,I_i}^{[h-1]}}\cdot d_i\right);
\end{eqnarray*}
\item[(ii)] for $j\neq i$, renormalize each $\psi_{i,j}^{[h]\ast}$ by multiplying  with $1-\psi_{i,\,I_i}^{[h]\ast}$.
\end{description}
\end{enumerate}
\rule[0.5ex]{\textwidth}{0.1mm}
}
\normalsize

For a more general introduction to copulas, see for instance \citet{NEL06} or \citet{GEN07}, as well as \citet{KIM07} for more specific issues about copulas fitting. 

In our experiments, we used a Gaussian copula to sample the new diagonal parameters, mainly because of its symmetric properties and its simplicity of calibration using a correlation matrix $\bf R$ \citep{MAR88}. Note that one has to consider and check up with great care the $p$ previously simulated matrices $\{\tilde{\bm{\psi}}^{[1]}, \dots, \tilde{\bm{\psi}}^{[p]}\}$ to make sure that a robust empirical estimator of $\bf R$ can be defined, in the sense that its Cholesky decomposition 
is numerically stable during the sampling process \citep{MAR88}. The {\it condition number} can be used to do so \citep{ELG02}. 
Conditionally on correlated sampled parameters, Dirichlet distributions appear necessary to get coherent instrumental sampling of remaining elements within each row vector $\bm\psi^{[h]\ast}_i$. 

\paragraph{Theoretical behavior} Despite the large amount of existing work aiming to simplify the conditions ensuring ergodicity and stationarity of the target distribution \citep{NOTT2005,ROBS07,ROBS09,ATCHADE2010}, theoretical descriptions of kernels based on Dirichlet products compounded with Gaussian copulas turn out to be 
technically complex, and their study deserves a specific work which remains outside the scope of this article. Since our primary aim is to assess the interest of exploring the correlations between the rows of $\bm\psi$, we adopt the simplest approach of  a {\it finite sampling scheme} when choosing $J$, as proposed by \citet{ROBS07}: given a  time $\tau<\infty$,  $J_{\Gamma_n}=J_{\Gamma_{\tau}}$ for any $n\geq \tau$. This approach is carried out in this paper at each sweep of the algorithm after a given mixing period, selecting the final $J_{\Gamma_{\tau}}$ as the basic product of Dirichlet's described hereinbefore. In substance, $\tau$ has the sense of an exploration time, and in practice  is selected as the minimum time between the time required for a fixed number of iterations and the time until the Brooks-Gelman RT is fulfilled.

Nonetheless, this explorative study fits into recent schemes shared by several authors, who tested copula-based methods to improve the efficiency of their sampling algorithms. In their seminal work on the optimization of the adaptation, \citet{HAARIO2001} considered Gaussian copula instrumental distributions calibrated over the full past of the chains. See \citet{AND08} for a review of this particular major field of adaptive MCMC.  \citet{STRID2010} used the sampling history to continuously calibrate a $t-$copula proposal distribution, in order to sample from dynamic stochastic equilibrium models.  Finally, \citet{CRAIU2011} used products of bivariate copulas to tune MCMC during an initialization period only, in the same spirit as the finite sampling approach used in the present paper.

\begin{figure}[h]
\centering
\includegraphics[height=8cm,width=10cm]{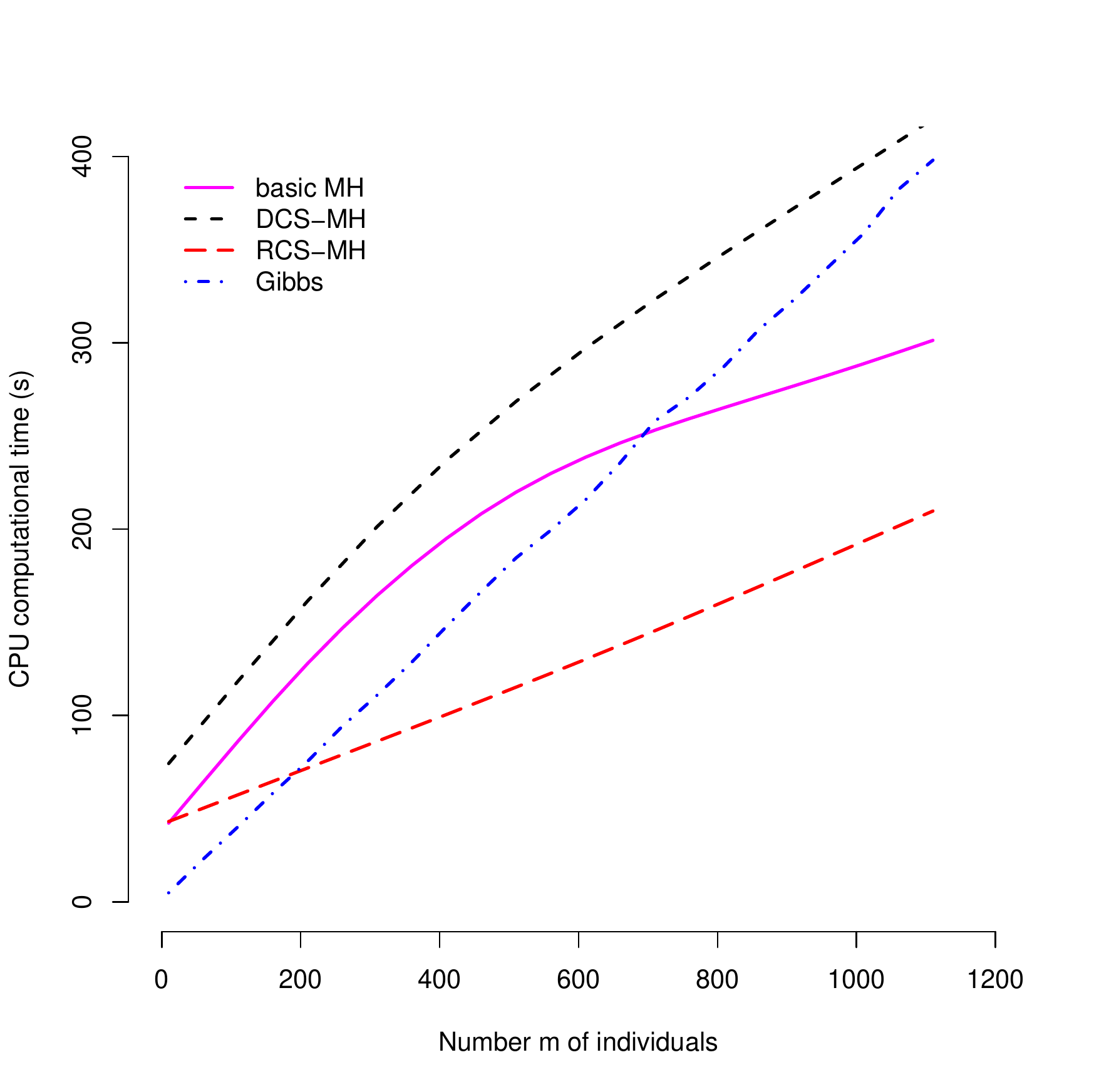}
\caption {Case study of \citet{LEE68}. Mean CPU time needed to reach quasi-stationarity as a function of the number $m$ of individuals (same simulations as Figure \ref{comparison-gibbs-mh-1}). With respect to the Figure \ref{comparison-gibbs-mh-1}, Gibbs and basic MH are also compared to DCS-MH and RCS-MH algorithms.}
\label{comparison.2}
\end{figure}

\paragraph{Illustration}
Continuing the four-dimensional simulated example from Section \ref{example1}, we applied the DCS-MH and RCS-MH methods with $p=30$, still augmenting the number $m$ of individuals and using three parallel chains per experiment. Parameters $d_i$ remain similarly sampled at each iteration. Results are smoothed over 50 similar runs of algorithms. 
The comparison of Gibbs and MH burn-in periods in Figure \ref{comparison.2}, in the sense of the Brooks-Gelman RT, illustrates the improvement yielded by RCS-MH. On the other hand, in this case DCS-MH performs worse than basic MH and even Gibbs sampler.
As we could expect, RCS-MH does clearly better than DCS-MH because of its widest exploration of the parameter space. RCS-MH strongly beats Gibbs even for relative low numbers of individuals.

The poor performance of DCS-MH is due to the computational cost of the selection of past matrices $\{\tilde{\bm{\psi}}^{[1]}, \dots, \tilde{\bm{\psi}}^{[p]}\}$ sufficiently different to allow for a robust Cholesky inversion. This cost clearly increases with the progression towards stationarity since sampled matrices become more and more similar and many among them must be rejected in the calibration task of the instrumental distribution. The RCS-MH algorithm suffers of course from the same defect, but the much better mixing counterbalances the increase of the computational cost, with respect to the basic MH algorithm, in a significant way. 

\section{Numerical experiments}\label{tests}

 This section deals with simulation studies to test the potentialities of our adaptive proposals to a wide class of transition matrices commonly encountered in reliability and risk assessment (RRA).  In RRA, it often occurs that 
the degradation of a system $\Sigma$ is described using $r$ separated states (for instance defined by a scale of crack sizes), ordered from minor defect to major failure (replacement cause). To be conservative, one may assume that potential repairs following a running failure are, at best, {\it as bad as old}, namely $\Sigma$ remains in the same state than before the failure. In other cases, one might assume these repairs bring actually more complications than real improvement (for instance if $\Sigma$ is old), so that $\Sigma$ is more deteriorated after the repair than before ({\it worth than old} repair). See \citet{BAS07} for more details about these notions. Under a stationarity assumption, the transition matrix $\bm\psi$ is necessarily upper triangular, with $\bm\psi_{r}=(0,\ldots,0,1)$. 

\paragraph{Simulation features}
In the following experiments, we test the potentialities of Gibbs and the three MH algorithms described hereinbefore (basic, DCS-MH and RCS-MH) as a function of $r$. We vary the dimension $r$ between 2 and $r_{\max}$ (in practice, we consider $r_{\max}=6$ to remain realistic).
To start with, we need a rule to sample realistic matrices with decreasing dimension:
\begin{enumerate}
\item  denote $\bm\psi^{(r)}$ a $r\times r$ upper triangular matrix. 
\item create $\bm\psi^{(r-1)}$ matrix as follows: for $i=1,\ldots,r-1$,
\begin{eqnarray*}
\psi^{(r-1)}_{i,j} & = & \psi^{(r)}_{i,j} \ \ \text{for $j=1,\ldots,r-2$} 
\end{eqnarray*}
and
\begin{eqnarray*}
\psi^{(r-1)}_{i,r-1} & = & \psi^{(r)}_{i,r-1} + \psi^{(r)}_{i,r}.
\end{eqnarray*}
\end{enumerate}
Doing so we automatically ensure that $\bm\psi^{(r-1)}_{r-1}=(0,\ldots,0,1)$. The rationale for this construction is obviously to increase the probability of a major failure event when simplifying the model. Thus we simply need to sample $\bm\psi^{(r_{\max})}$ to get all other matrices considered for simulation tests. Pursuing our wish of realism, we assume that {\it worth than old} repairs are less probable than {\it as bad as old} ones. Therefore, for $i=1,\ldots,r_{\max}-2$ and $k=1,\ldots,r_{\max}-i-1$, we 
assume in the sampling:
\begin{eqnarray*}
\psi^{(r_{\max})}_{i,i} \ > \ \psi^{(r_{\max})}_{i,i+k}  >  \sum\limits_{p=k+1}^{r_{\max}-i} \psi^{(r_{\max})}_{i,i+p},
\end{eqnarray*} 
and especially for $i=r_{\max}-1$,
$\psi^{(r_{\max})}_{r_{\max}-1,r_{\max}-1}  >  \psi^{(r_{\max})}_{r_{\max}-1,r_{\max}}$
to ensure a constant decreasing of values $\psi^{(r)}_{i,i},\psi^{(r)}_{i,i+1},\ldots,\psi^{(r)}_{i,r}$ for any $r\leq r_{\max}$. Finally, we selected matrices $\bm\psi^{(r_{\max})}$ for which:
\begin{eqnarray*}
\psi^{(r_{\max})}_{i,i}  \leq   \psi^{(r_{\max})}_{i+1,i+1}.
\end{eqnarray*}   

This models the following case: the closer to a major failure state, the better (the more cautious) the repair. Notice that we do not take into account any of our simulation constraints in the following estimation procedures, except the presence of zeros beneath the diagonal of $\bm\psi$ (by reducing the length of Dirichlet distributed vectors in the instrumental sampling).
We consider it as a minimal knowledge assumable in real case-studies (e.g. Section~\ref{complete.exemple}). 
Finally, per simulated matrix, a complete sequence for $m=1000$ individuals was generated for $T=20$ observation times. 
As we are always in the particular case of ``a single observation per individual'', only one observation is retained in each sequence for the inference exercise.

\paragraph{Estimation} As in Section \ref{example1}, each experiment for a given $r\in[3,r_{\max}=6]$ consists in running  three parallel chains for each method and monitoring them using the Brooks-Gelman statistic.  Relative Euclidian errors on posterior means of matrix components (computed using 1000 iterations after a burn-in period determined by the Brooks-Gelman RT) are fixed at most at 5\%, involving preliminary tests for fixing the total number of iterations. 
Again, parameters $d_i$ are sampled uniformly in $[100,2500]$. 
Finally, each experiment is repeated 100 times to average the results (each time a new family of matrices $\bm\psi^{(r_{\max})},\ldots,\bm\psi^{(3)}$ being simulated). 

\begin{figure}[h]
\centering
\includegraphics[height=9cm,width=9.5cm]{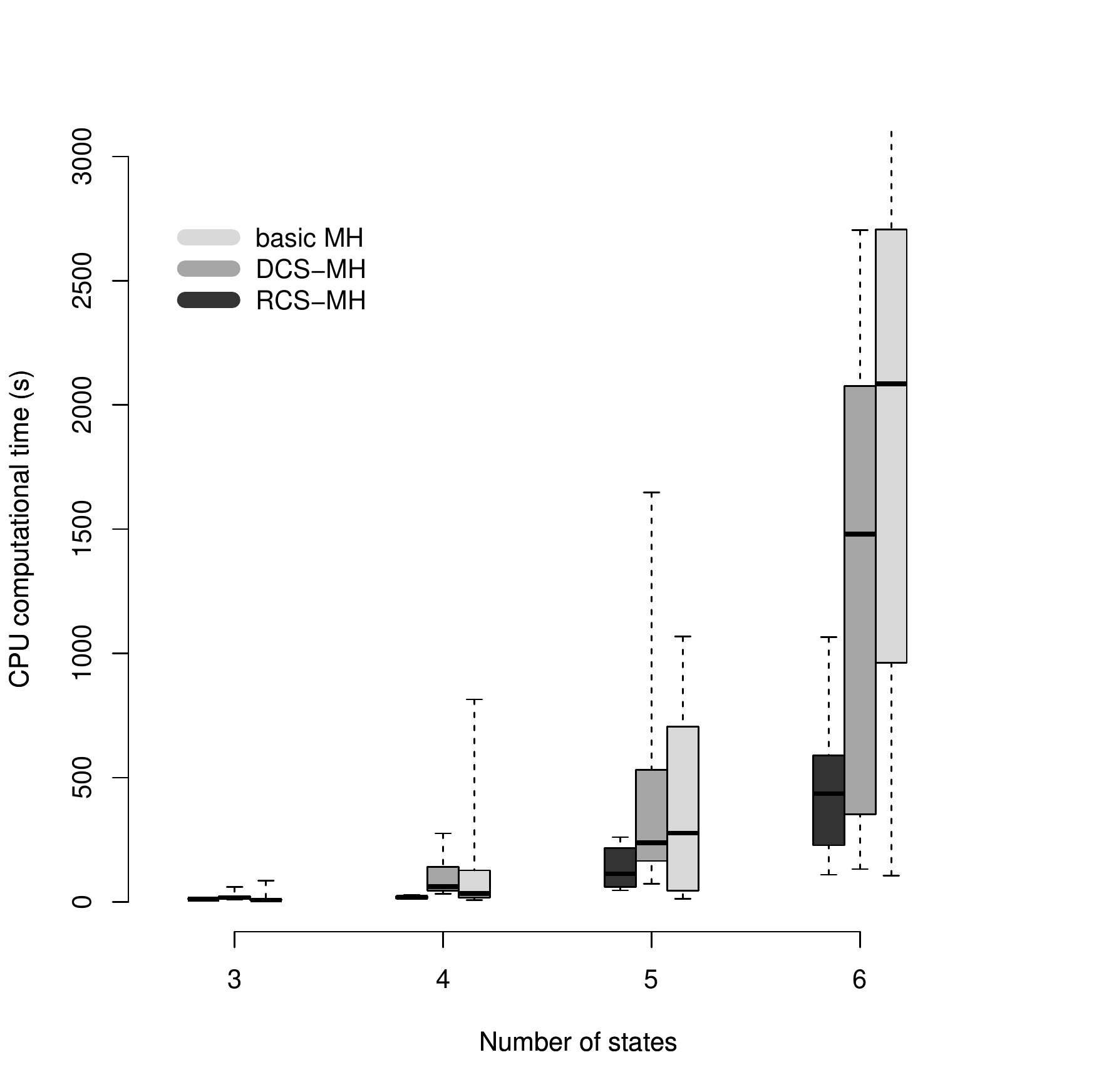}
\caption{RRA case study. Boxplots of CPU times needed to reach quasi-stationarity as a function of the dimension $r$. Half lines indicate median and bounds indicate most extreme values.  Data have been generated by upper-triangular transition matrices.}
\label{box}
\end{figure}

\begin{figure}[h]
\centering
\includegraphics[height=9cm,width=8.6cm]{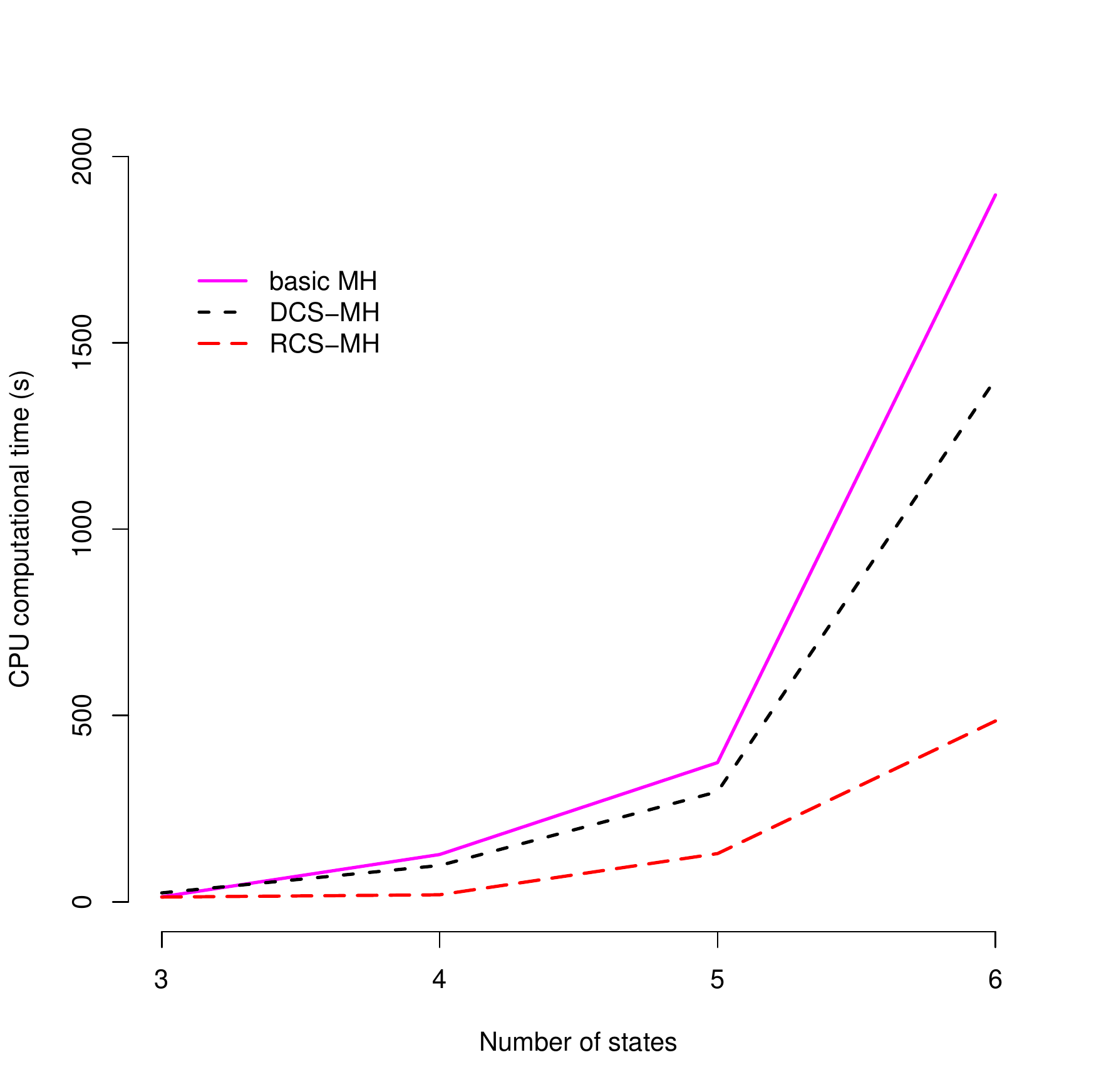}
\caption{RRA case study. Mean CPU times needed to reach quasi-stationarity.}
\label{meancpu}
\end{figure}

\paragraph{Results} Boxplots and mean CPU times before quasi-stationarity (in the sense of the Brooks-Gelman RT) are plotted in Figures \ref{box} and \ref{meancpu}. Results obtained on the simulated example from Section \ref{example1} can be generalized: RCS-MH provides for all dimensions a significant improvement in mixing. Similar results have been obtained when carrying out an empirical approach to calibrate the mean acceptance rate to a standard nominal value of 50\% then 25\%. 

 \section{Discussion}\label{discussion}
 
 \subsection{Main ideas and results}
 
This article first aims to provide a general review and technical advises about the Bayesian estimation of finite-state transition matrices $\bm\psi$ in discrete Markovian models under various missing data schemes, which appear to be of particular interest in several domains, especially in engineering. Actually, reliability practitioners may frequently deal with classes of upper-triangular transition matrices that have been chosen for most of the experiments presented here. Depending on the nature of available data, the practitioner may have to choose between Gibbs or Metropolis-Hastings (MH) algorithms. The time-consuming features of these algorithms, depending on the size of missing data and the dimension of the problem, appear as limiting factors in practice. Therefore, the second part of this article focuses on a first exploration of two adaptive mechanisms (DCS-MH and RCS-MH) likely to accelerate the MH algorithms. 

Numerical experiments have highlighted, on this specific class of examples, that using instrumental distributions based on Gaussian copulas to account for the correlations between the rows of $\bm\psi$ yields a better mixing of the chains, implying a significant reduction of the computational cost.
The gap with basic MH strategies, based on the independent sampling of the rows of $\bm\psi$, increases with the number $m$ of individuals or the number $r$ of states. The simplicity of the approaches proposed here lets us think that any practitioner dealing with aggregate data could easily implement the DCS-MH and RCS-MH mechanisms and reduce the computational time. 

\begin{figure}[hbt]
\centering
\includegraphics[height=9cm,width=9.5cm,angle=90]{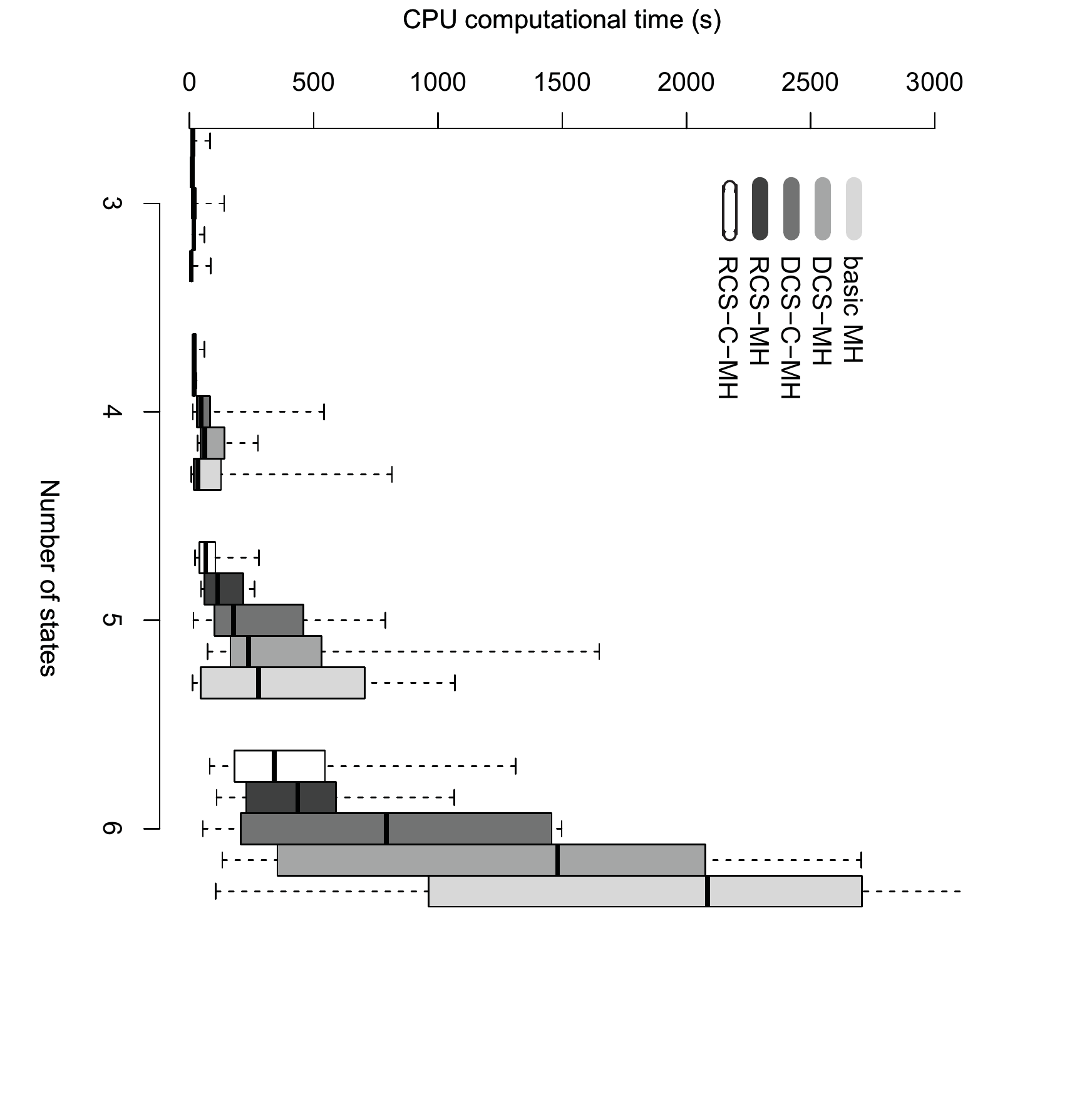}
\caption{RRA case study. Boxplots of CPU times needed to reach quasi-stationarity as a function of the dimension $r$. Half lines indicate median and bounds indicate most extreme values.  Data have been generated by upper-triangular transition matrices. The DCS-C-MH and RCS-C-MH abbrevations indicate two ``coarse" versions of the DCS-MH and RCS-MH mechanisms.}
\label{CPU-final}
\end{figure}

Supplementary experiments have highlighted that the CPU time can be still diminished by using two ``coarse" versions of the DCS-MH and RCS-MH mechanisms. They consist in estimating the copula parameter ${\bf R}$ directly from the Pearson correlation of the matrix elements, namely removing the step $2.{\bf (ii)}$ in each mechanism. These coarse approaches (we call them DCS-C-MH and RCS-C-MH) have been be compared to the previous ones in Figure~\ref{CPU-final}. Here, the difference in CPU time is mainly due to the cost of empirical inversions in the DCS-MH and RCS-MH methods.

The adaptive schemes proposed here (especially the most powerful RCS-MH and RCS-C-MH), which remain only empirically studied, deserve a more specific study from both theoretical and applied viewpoints. This point is more widely discussed in the following subsection.

As a take-home message, in the most general case of incomplete data problems with several observations per individual, the Gibbs sampler based on the data augmentation technique seems to be the only possible alternative. In the particular case of a single observation per individual, the adaptive MH algorithms (and especially RCS-MH) are valid alternatives to the Gibbs sampler if the number of individuals is greater than a few hundred, say 200, and the number of states is greater than three. In low dimensional problems (two or three) the practical interest of adaptive MH methods, with respect to the simpler Gibbs sampler, is less obvious.

\subsection{Directions of further research}
 
The adaptation processes proposed here remain empirical, and theoretical studies are needed to build copula-based strategies ensuring the ergodicity and the stationarity of the chains less crudely than imposing a finite adaptation time, based on principles initiated by \citet{ROBS07} and \citet{AND06}. Indeed, fully adaptive MCMC should be build on {\it infinite} adaptations which continuously modify the choice of the transition kernel using the past values of $\bm\psi$ along the chains, quasi-stationarity occurring when these kernel modifications become imperceptible.  
These adaptations should be led on both correlated and marginal features of the matrix elements. To this first aim, future studies could focus on the mechanism of state permutation, inspired by similar ones carried out in the framework of variable selection \citep{NOTT2005}, and on removing the strong assumption made by using a Gaussian copula to model correlations within the elements of $\bm\psi$. This choice can appear oversimplified since it does not take into account possible correlations between extreme values in the instrumental distribution of $\bm\psi$.
Therefore a copula selection procedure should be carried out at different times of the adaptive chain, for instance  using frequentist tests (e.g. Cramer-von Mises), based on distances between estimated and simulated copulas  \citep{GEN06,NIK08} or Bayesian posterior odds \citep{HUA06}. As those procedures remain time-consuming in dimensions $r\geq 2$, this approach was not implemented here in this exploratory work. 

Furthermore, it is necessary that such more sophisticated adaptive Metropolis-Hastings algorithms be compared in practice to refined Gibbs algorithms, evoked at the end of Section~\ref{example1}, that could benefit from the stick-breaking properties of Dirichlet distributions. 
 
Another point of interest could be the adaptation of the methods reviewed here to the case of non stationary Markov chains. A simple way for doing this could be to stratify the data on the time $t$ or on groups of values of $t$ \citep{URA75,SEN99}. The use of logit or proportional odds models \citep{COL05,GRI11} to include also the effect of additional covariates is another perspective for this work.
 

\section*{Acknowledgments}

The authors gratefully thank Prof. Eric Parent (AgroParisTech), as well as Drs. Merlin Keller (EDF R\&D) and Sophie Ancelet (IRSN) for many discussions and advices. Many other thanks are addressed to the Associate Editor and two anonymous reviewers who greatly helped to improve the clarity and refine the scope of this article. This work was partially supported by the French Ministry of Economy in the context of the CSDL (\textit{Complex Systems Design Lab}) project of the Business Cluster System@tic Paris-R\'{e}gion.


\appendix
\section{Proof of expression (\ref{likelihood.once.observed})}\label{app:proof}
The probability of a given sequence of $T+1$ states ($t$ from 0 to $T$) with only one observed state at $t=t_k$ can be written as the sum of $T$ sums fits into each other (with index from $1$ to $r$), one for each unobserved state:
\begin{eqnarray*}
\mathbbm{P}(\bullet,...,\bullet,s_j ,\bullet,...,\bullet) \ \ \ \text{is equal to} 
\end{eqnarray*}
\begin{eqnarray*}
\sum\nolimits_{i_{0}}p_{i_{0}}(0)\bigg[ \sum\nolimits_{i_{1}}\psi _{i_{0},i_{1}}\bigg[\sum\nolimits_{i_{2}}\psi _{i_{1},i_{2}}...
  \bigg[  \sum\nolimits_{i_{t_{k}-1}}\psi_{i_{t_{k}-2},i_{t_{k}-1}}\psi _{i_{t_{k}-1},j}\bigg[ \sum\nolimits_{i_{t_{k}}}\psi_{j,i_{t_{k}}}...  \bigg[ \sum\nolimits_{i_{T}}\psi _{i_{T-1},i_{T}}\bigg]... \bigg] \bigg] ...\bigg] \bigg] 
\end{eqnarray*}
In the expression above the sum of the first $t_k$ sums (indices from $i_0$ to $i_{t_{k}-1}$) is the unconditional probability $p_j(t_k)$ for the system to be in state $j$ at time $t_k$. That can be showed by developing the recursive formula (\ref{probabilities.0}): 
\begin{eqnarray*}
p_j (t_k ) & = & \sum\nolimits_{g_1 } {p_{g_1 } (t_k  - 1) \cdot \psi _{g_1 ,j} } \\
 & = & \sum\nolimits_{g_1 } {\sum\nolimits_{g_2 } {p_{g_2 } (t_k  - 2) \cdot \psi _{g_2 ,g_1 } }  \cdot \psi _{g_1 ,j}}  \\ 
& = &  \sum\nolimits_{g_1 } {\sum\nolimits_{g_2 } {\sum\nolimits_{g_3 } {p_{g_3 } (t_k  - 3)}  \cdot \psi _{g_3 ,g_2 }  \cdot \psi _{g_2 ,g_1 } }  \cdot \psi _{g_1 j}  }  \\
& = & \ldots 
\end{eqnarray*}
and renaming the index $g_1, g_2, g_3, \ldots$ as $i_{t_{k}-1}, i_{t_{k}-2}, i_{t_{k}-3},\ldots$. The sum of the remaining sums in  (\ref{one.step.trans}) 
is one as $\sum_j {\psi _{i,j}  = 1}$. The probability of the sequence is then $p_j(t_k)$. Thus, the likelihood of $m$ incomplete sequences where each individual is observed only once can be written:
\begin{eqnarray*}
\prod\limits_{k = 1}^m {\prod\limits_{j = 1}^r {\prod\limits_{t = 0}^T {p_j (t)^{\mathbbm{1}_{\{ {t = t_k ,y(k,t) = s_j } \} }} } } }  =  \prod\limits_{j = 1}^r {\prod\limits_{t = 0}^T {p_j (t)^{\sum\nolimits_k {\mathbbm{1}_{\{ {t = t_k ,y(k,t) = s_j } \} }} } } } 
\end{eqnarray*}
which is, under the hypothesis that probabilities $p_j(0)$ are know, the expression (\ref{likelihood.once.observed}) up to constant of proportionality.
\section{Gibbs sampler for MNAR data when missingness only depends on the actual state}\label{app:GibbsMNAR}
First initialize the algorithm by arbitrarily completing state sequences. Then at each step $h=1, 2,\ldots$, perform the following two-step procedure:
\begin{enumerate}
\item parameters estimation:
\begin{eqnarray*}
{\bm{\psi }}_i^{[h]} {\rm{|}}{\bm y}^{[h-1]} & \sim&  \Dir\left(\gamma _{i1}  + w_{i,1}^{[h-1]} ,...,\gamma _{ir}  + w_{1,r}^{[h-1]}\right)
\end{eqnarray*}
and
\begin{eqnarray*}
\eta_i^{[h]} {\rm{|}}{\bm y^{[h-1]}} & \sim & \Be\left(\alpha _{i} + a_{i}^{[h-1]} , \beta _{i} + b_{i}^{[h-1]} \right),\\
\end{eqnarray*}
where $a_i^{[h-1]} = \sum\limits_{t = 1}^T {\sum\limits_{k = 1}^m {\mathbbm{1}_{ \left\{y_{(k,t)}^{[h-1]} = s_i ,x_{(k,t)} = 1 \right\} } }}$, $b_i^{[h-1]} = \sum\limits_{t = 1}^T {\sum\limits_{k = 1}^m {\mathbbm{1}_{ \left\{y_{(k,t)}^{[h-1]} = s_i ,x_{(k,t)} = 0 \right\} } }}$ and $w_{i,j}^{[h-1]}$ are the same as in Section \ref{review.incomplete}.
\item data augmentation: drawing $z^{[h]}_{\texttt{mis}(k,t)}$ conditional on the following probabilities:
\begin{eqnarray*}
\mathbbm{P} \left( {y^{[h]}_{(k,1)} = s_j {\rm{|}}y^{[h - 1]}_{(k,2)}=s_i,{\bm{\psi }}^{[h]} , {\bm{\eta }}^{[h]} } \right) & \propto & \eta_j^{[h]} \cdot \psi _{j,i}^{[h]}, \ \ \text{for $t=1$},
\end{eqnarray*}
\begin{eqnarray*}
\mathbbm{P} \left( {y^{[h]}_{(k,T)} = s_j {\rm{|}}y^{[h]}_{(k,T - 1)}=s_i,{\bm{\psi }}^{[h]} , {\bm{\eta }}^{[h]} } \right) & \propto & \eta_j^{[h]} \cdot {\psi _{i,j}^{[h]}}, \ \ \text{for $t=T$}
\end{eqnarray*}
and
\begin{eqnarray*}
\mathbbm{P} \left( {y^{[h]}_{(k,t)} = s_j {\rm{|}}y^{[h]}_{(k,t - 1)}=s_{i_1},y^{[h - 1]}_{(k,t + 1)}=s_{i_2},{\bm{\psi }}^{[h]}, {\bm{\eta }}^{[h]} } \right) & \propto & \eta_j^{[h]} \cdot \psi _{i_1,j}^{[h]}  \cdot \psi _{j,i_2}^{[h]}, \ \ \text{otherwise}.
\end{eqnarray*}
\end{enumerate}



\bibliographystyle{elsarticle-harv}


\end{document}